\documentclass[10pt,preprint]{aastex6}

\usepackage{color,xspace}
\usepackage{epsfig}

\newcommand{\cii}{[\ion{C}{2}]\xspace}

\shorttitle{Galaxy Formation Through Filamentary Accretion at \lowercase{$z=6.1$}}

\shortauthors{Jones et al.}
 
\begin{document}

\title{Galaxy Formation Through Filamentary Accretion at \lowercase{$z=6.1$}}

\author{G. C. Jones\altaffilmark{1,2},
C. J. Willott\altaffilmark{3},
C. L. Carilli\altaffilmark{1,4},
A. Ferrara\altaffilmark{5,6},
R. Wang\altaffilmark{7},
J. Wagg\altaffilmark{8}
}

\altaffiltext{1}{National Radio Astronomy Observatory, 1003 Lopezville Road, Socorro, NM 87801, USA; gcjones@nrao.edu} 
\altaffiltext{2}{Physics Department, New Mexico Institute of Mining and Technology, 801 Leroy Pl, Socorro, NM 87801, USA} 
\altaffiltext{3}{NRC Herzberg, 5071 West Saanich Rd, Victoria, BC V9E 2E7, Canada} 
\altaffiltext{4}{Astrophysics Group, Cavendish Laboratory, JJ Thomson Avenue, Cambridge CB3 0HE, UK}
\altaffiltext{5}{Scuola Normale Superiore, Piazza dei Cavalieri 7, I-56126 Pisa, Italy}
\altaffiltext{6}{Kavli IPMU, The University of Tokyo, 5-1-5 Kashiwanoha, Kashiwa 277-8583, Japan}
\altaffiltext{7}{Kavli Institute of Astronomy and Astrophysics at Peking University, No.5 Yiheyuan Road, Haidian District, Beijing, 100871, China}
\altaffiltext{8}{SKA Organization, Lower Withington Macclesfield, Cheshire SK11 9DL, UK}

\begin{abstract}

  We present ALMA observations of the dust continuum and [\ion{C}{2}]
  158$\mu$m line emission from the $z=6.0695$ Lyman Break Galaxy
  WMH5. These observations at $0.3''$ spatial resolution show a compact
  ($\sim3$\,kpc) main galaxy in dust and \cii emission, with a `tail'
  of emission extending to the east by about 5\,kpc (in projection).
  The \cii tail is comprised predominantly of two distinct
  sub-components in velocity, separated from the core by $\sim100$ and $250$\,km\,s\,$^{-1}$, with narrow intrinsic widths of about
  80\,km\,s$^{-1}$, which we call `sub-galaxies'.  The sub-galaxies
  themselves are extended east-west by about $3$\,kpc in individual
  channel images. The \cii tail joins smoothly into the main galaxy
  velocity field.  
  The \cii line to continuum ratios are comparable for
  the main and sub-galaxy positions, within a factor 2. In addition, these ratios are comparable to $z\sim5.5$ LBGs.
  We conjecture
  that the WMH5 system represents the early formation of a galaxy
  through the accretion of smaller satellite galaxies, embedded in a
  smoother gas distribution, along a possibly filamentary
  structure. The results are consistent with current cosmological
  simulations of early galaxy formation, and support the idea of very
  early enrichment with dust and heavy elements of the accreting
  material.

\end{abstract}


\section{Introduction}

The formation of the first galaxies in the Universe, those at $z > 6$,
or within 1\,Gyr of the Big Bang, is a forefront question in modern
astronomy. How these galaxies accrete the gas that drives early star
formation, and the relative dust attenuation,
remains highly uncertain. These issues have attained new pertinence as
constraints on the evolution of the neutral fraction of the
intergalactic medium have solidified. Current observational
constraints based on e.g. the cosmic microwave background (CMB), the Gunn-Peterson effect, and the 
effect of the neutral intergalactic medium (IGM) on
Ly$\alpha$ emission from early galaxies, suggest that the Universe was
highly ionized at $z < 6$, and significantly neutral at $z > 7$, with
a plausible neutral fraction of 50\% at $z \sim 7.5$
(\citealt{fan06a,rob15,ota17,grei17}).  This evolution in the cosmic neutral
fraction suggests substantial star formation at redshifts $z \sim 6$
to 8, if the IGM is reionized by star forming galaxies
(\citealt{fan06b}).

The Atacama Large Millimeter/submillimeter Array (ALMA) has opened a new window on the
early Universe, with the sensitivity and resolution to perform
detailed studies of the dust and cool gas in the earliest galaxies.
Of particular interest are observations of the \cii 158\,$\mu$m line at
$z > 6$.  This line is typically the brightest emission line from star
forming galaxies at meter through FIR wavelengths, tracing both the
cool and warm interstellar medium (ISM) in galaxies (\textit{e.g.},
\citealt{pine13,cariw13}). Observation of dust in the most distant
galaxies provides key constraints on the extinction properties, while
the \cii line provides a means of tracing gas kinematics and obtaining information on the
structure of the ISM. These
quantities relate to the star formation rate of the galaxy.

Cosmological simulations of early galaxy formation have shown that the
\cii line can be used not only to study the ISM of these sources
(\citealt{vall13,vall15,pall17}), but it might also probe the overall
galaxy dynamics, including supernova-driven outflows \citep{gall16}.
In the local universe a well-established \cii-star formation rate (SFR) relation holds for
a wide range of galaxy types, from metal poor dwarf galaxies,
to starbursts, ultra-luminous infrared galaxies, and active galactic nucleus (AGN) hosting
galaxies (\citealt{delo14,pine14,herr15}). However, many of the
high-$z$ Lyman-Break Galaxies (LBGs) and Lyman-$\alpha$ Emitting
Galaxies (LAEs) deviate from this relation for yet unclear reasons,
likely related to their different metallicity, ionization state, or
more efficient supernova feedback \citep{capa15}.

The galaxy WMH5 at $z = 6.0695$ is a useful test-bed to investigate
detailed processes in the formation of early galaxies. This galaxy was
discovered in the LBG search of \citet{will13} as
one of the most luminous LBGs at $z > 6$ (near-IR magnitude AB $\sim
24$), with detected Ly$\alpha$ emission.
\citet{will15} discovered both thermal emission from dust,
and \cii line emission, from this galaxy in Cycle 2 ALMA observations at $0.5''$ resolution. They presented a detailed analysis
of the integrated spectral energy distribution (SED) of the galaxy from the optical through submm. 
The inferred SFR values are between 33 and 66\,$M_{\odot}$\,year$^{-1}$, depending
 on the tracer, and the stellar mass is $2.3\times
10^{10}M_\odot$. The overall SED has dust attenuation
of $E(B-V) = 0.05$. However, the main component of the galaxy seen 
in the dust and \cii is spatially offset $0.4''$ west of 
the bright NIR component. The \cii at the NIR position
is also offset significantly in velocity relative to the main \cii
component ($\sim 200$km\,s$^{-1}$). \citet{will15} suggest
a merging galaxy system, however, the spatial resolution and sensitivity
of these early ALMA observations were insufficient to determine the detailed
structure of this complex system.

In this paper, we present subarcsecond angular resolution and high sensitivity
observations of the \cii and dust emission from WMH5 using ALMA in
Cycle 3.  These observations present a clear picture of the processes
occurring in WMH5. We  assume
($\Omega_{\Lambda}$,$\Omega_m$,h)=(0.692,0.308,0.678) \citep{plan16}
throughout. At this distance, 1 arcsecond corresponds to 5.81\,proper kpc
at the redshift of WMH5 ($z$=6.0695; \citealt{will15}).

\section{Observations}

The \cii 158\,$\mu$m and thermal dust emission from WMH5 were
observed with ALMA in Cycle 3 in configuration C40-3 between 2016 August 3 and 7. 
Between 40 and 43 antennas were used for each observation. The
total time was $\sim6.5$\,hours, out of which 4\,hours were spent on
source. A total bandwidth of 7.5\,GHz was used, split into four 2\,GHz
sub-bands spanning 253 to 257\,GHz and 268 to 272\,GHz (ALMA band 6). One sub-band was centered on the
\cii emission, redshifted to $268.836$\,GHz, while the other
three were used to measure the rest frame $\sim160\,\mu$m continuum
emission.  The Cycle 3 data were calibrated using the standard
calibration script prepared by the Joint ALMA Observatory staff.

Using only the Cycle 3 data results in detections of both line 
and continuum emission. However, to improve the quality of the data,
we combined these data with the Cycle 2 observations of \citet{will15}. 
In short, they observed in June 2014 with 29-32 antennas, 7.5\,GHz of 
bandwidth, and had 95 minutes of on-source integration time.

An issue arose while combining the two datasets, based on the
relative weights of the \textit{uv} data. 
The definition of the weight factor as a function of channel width, 
integration time, and system temperature changed on September 4, 2014, when 
CASA updated from version 4.2.1 to 4.2.2\footnote{https://casa.nrao.edu/Memos/CASA-data-weights.pdf}. 
Since the Cycle 2 observations were made before this time and the Cycle 3 
observations after, the average weights differed by multiple orders of 
magnitude (i.e., $w\sim0.025$ for Cycle 2 and $w\sim20$ for Cycle 3). 
When using the CASA task \textit{clean} to combine these two datasets, 
this caused the Cycle 3 data to completely dominate. 

Two methods were tested to correct this. The first set all weights equal to one,
so that all data are given equal weight. The other used the CASA task 
\textit{statwt}, which resets weights based on the intrinsic scatter of the data. 
We compared images created using these approaches with the image made with the original weights. Both 
re-weighting methods resulted in images with lower rms noise levels than the 
original image, and similar channel maps. Since the \textit{statwt} 
image had the lowest rms noise level and had weights based on properties 
of the data, we will use this weighting method for the following analysis.

While different visibility weighting schemes were applied to the data
(\textit{i.e.}, Briggs weighting with robust parameters of 2.0, 1.5,
1.0), the best results were found for natural (robust 2.0)
weighting. The synthesized beam was $0.30''\times0.25''$, with major
axis position angle = $74^{\circ}$. Using the three sub-bands that were not
centered on the \cii line, we created a continuum image with
an RMS noise level of $9\,\mu$Jy\,beam$^{-1}$. The remaining sub-band was
used to create a line emission cube with $15.6\,\mathrm{MHz}\sim17$\,km\,s$^{-1}$ 
channels and an RMS noise level of 0.1\,mJy\,beam$^{-1}$.

\section{Results}

\begin{figure}
\centering 
\includegraphics[scale=0.8]{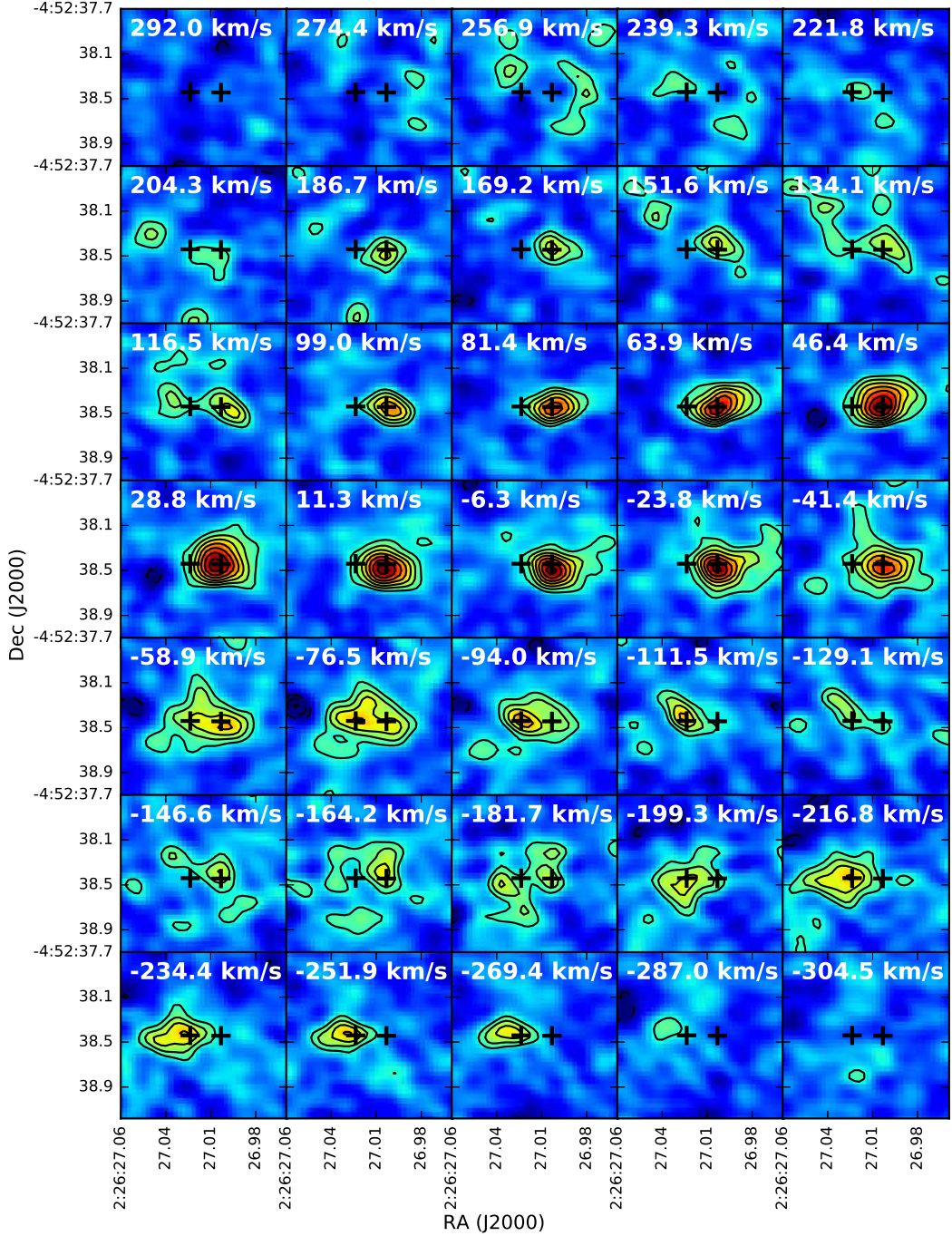}
\caption{Channel maps of \cii emission from WMH5. Contours 
begin at $\pm2\sigma$, where $1\sigma= 0.1\,m$Jy\,beam$^{-1}$, and are 
in steps of $1\sigma$. Crosses show locations of core and tail. 
Each channel spans 15\,MHz, so the entire displayed range is 292 to 
-305\,km\,s$^{-1}$, or 268.575 to 269.110\,GHz. The synthesized beam 
is $0.30''\times0.25''$, with major axis position angle = $74^{\circ}$.
North is up and east is to the left.}
\label{fig:channels}
\end{figure}

Figure \ref{fig:channels} shows the \cii channel images. We see emission from
$\sim+260$ to $-270$\,km\,s$^{-1}$, mostly going from west to east with increasing frequency (decreasing velocity). 
We also note that, in addition to this higher level emission, there is substantial (i.e., $\ge 2\sigma$) emission present to the east in select channels from $\sim +250$ to $+110$\,km\,s$^{-1}$.
We use these images and
those that follow to define two positions. The centers of 2-dimensional Gaussian fits to a map using only channels where 
emission is concentrated in the west, named the ``core'' (2h26m27.0025s -$4^{\circ}$52$'$38.4042$''$), 
and of the map made using the channels
where line emission extends to the east, named the ``tail'' (2h26m27.0208s -4$^{\circ}$52$'$38.4007$''$). In this
and subsequent analyses, we adopt zero velocity as \cii emission
redshifted to $z=6.0695$, consistent with \citet{will15}.

The channel images in Figure \ref{fig:channels} show that most of the
emission is concentrated in the core, but there are two distinct knots
of emission in the tail to the east at $\sim-110$ and $\sim-250$\,km
s$^{-1}$. These knots are more elongated east-west, extending EW by about $0.6''$ in individual
channels. 

\begin{figure}
\centering 
\includegraphics[scale=0.5]{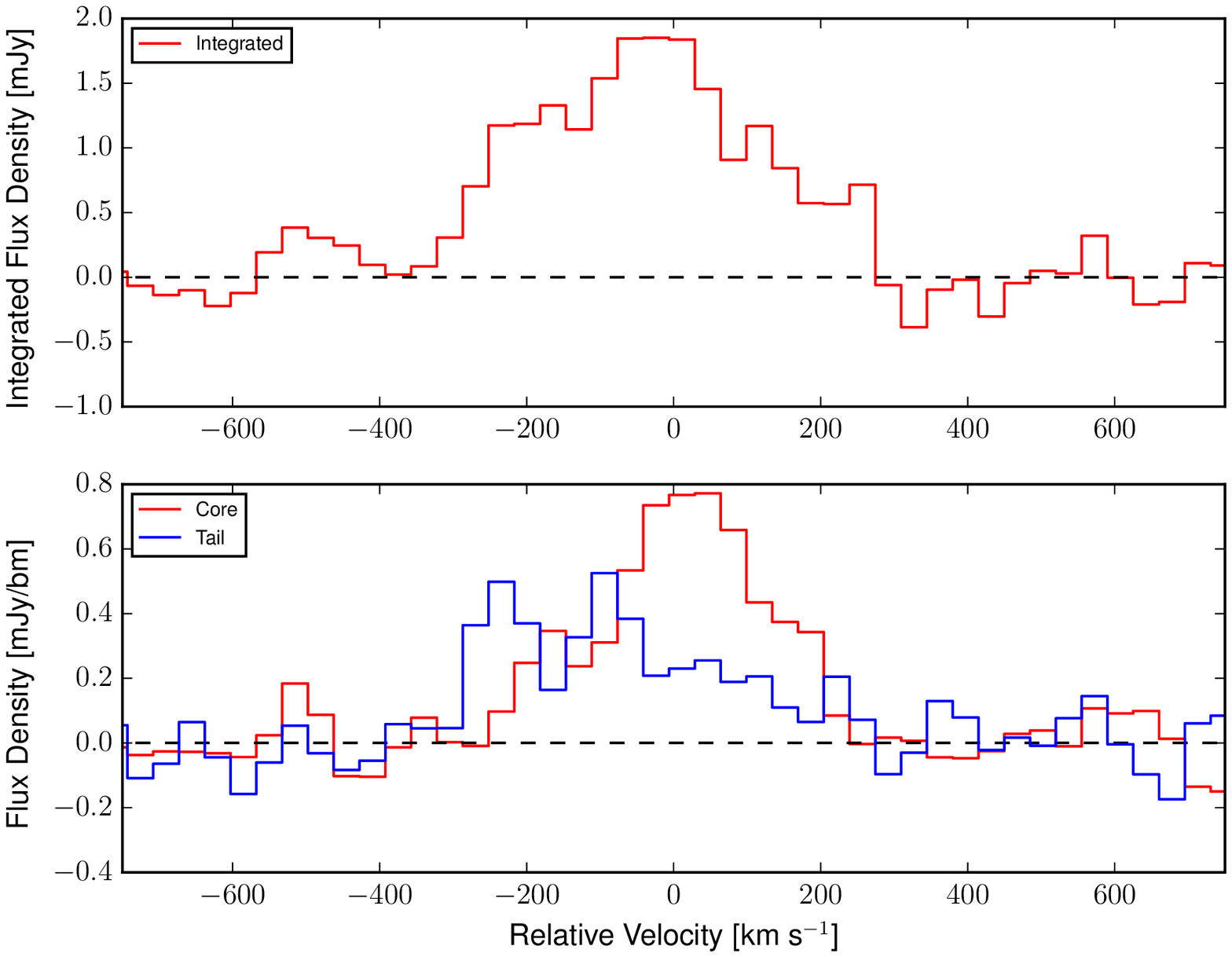} 
\caption{TOP: Integrated spectrum of full \cii emitting area. BOTTOM: point spectra at positions of the core (red) and tail (blue).}
\label{fig:integratedcii}
\end{figure}

\begin{figure}
\centering 
\includegraphics[scale=0.4]{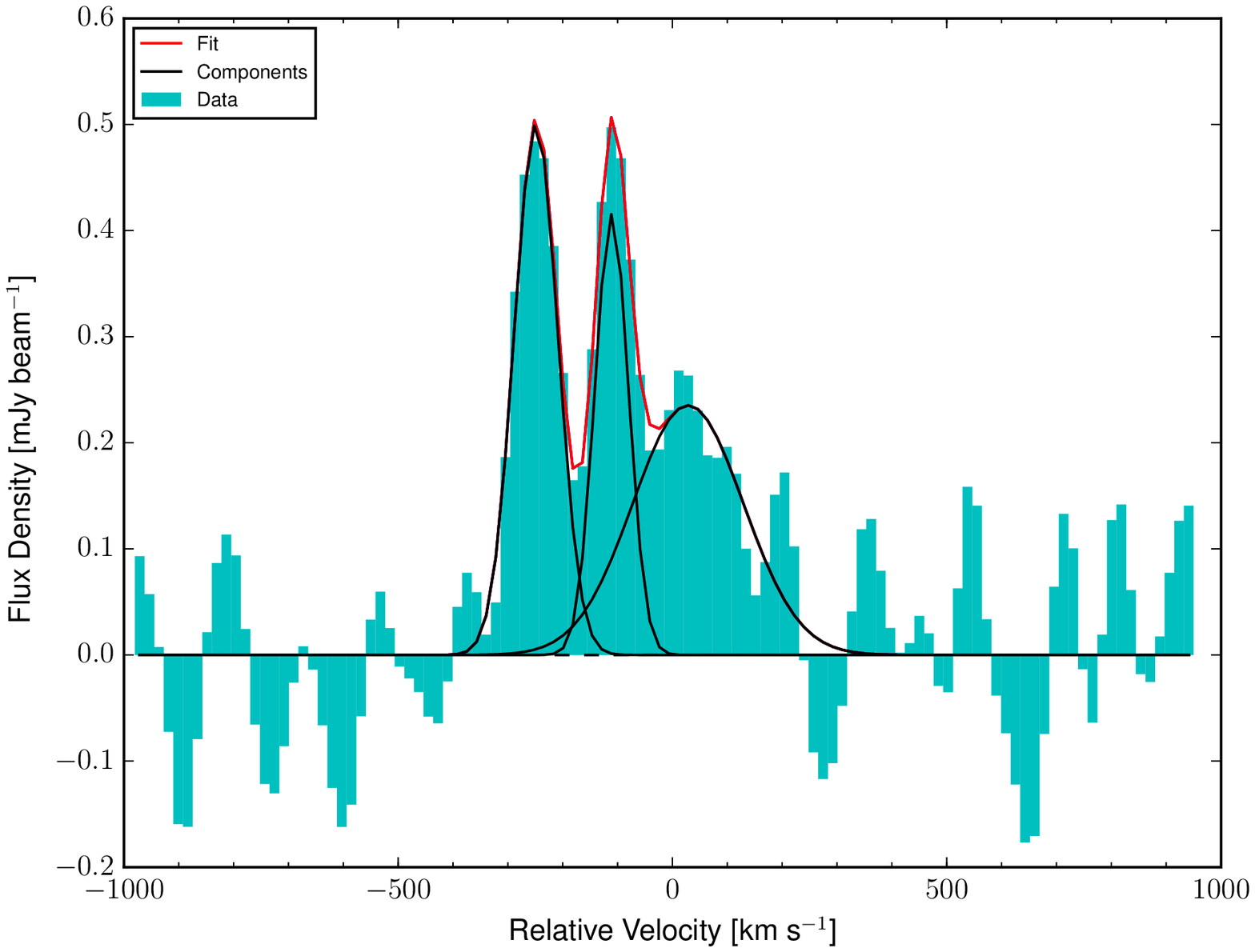}
\caption{Results of fitting three Gaussians to the spectrum at the ``tail'' position. 
See text for details.}
\label{speccomps}
\end{figure}

Figure \ref{fig:integratedcii} shows the integrated \cii
spectrum of the system and the spectra taken at the locations of
the core and tail, over a single spatial pixel in the cleaned image. While the core is well fit by a single Gaussian,
the tail is split into three different contributions: two narrow
peaks, which we call SG-1 (or Sub-Galaxy-1) and SG-2, and a spectrally
broad component, which we name ``Diffuse''. We performed Gaussian fitting to the apparent three
components of the tail. Figure \ref{speccomps} shows the resulting
fits for the tail. Details of the Gaussian fits for the core and tail are given in
Table \ref{north1}.

\begin{deluxetable}{lcccc} 
\tablecolumns{5}
\tablewidth{0pt}
\tabletypesize{\scriptsize}
\tablecaption{One-dimensional Gaussian fits to \cii components \label{north1}}
\tablehead{
Source  & Line Peak & Velocity & FWHM  & S$\Delta$v$_{\rm [CII]}$\\
 & [mJy\,beam$^{-1}$] & [km\,s$^{-1}$] & [km\,s$^{-1}$] & [Jy\,beam$^{-1}$km\,s$^{-1}$] }
\startdata
Core      & $0.76\pm0.03$   & $-1\pm5$           & $270\pm13$      & $0.22\pm0.01$ \\
SG-1     & $0.50\pm0.05$   & $-249\pm5$         & $94\pm12$     & $0.05\pm0.01$ \\
SG-2     & $0.4\pm0.1$    & $-111\pm5$         & $72\pm18$     &  $0.03\pm0.02$ \\
Diffuse     & $0.24\pm0.03$   & $28\pm34$            & $238\pm74$    &  $0.06\pm0.03$ \\ \hline
\enddata
\end{deluxetable}

The spectrally broad ``Diffuse'' component exhibits a similar centroid velocity as the core emission, suggesting that the diffuse emission is
the spatial edge of the core emission. Alternatively, the Diffuse emission may originate from numerous discrete, spectrally narrow components that we cannot separate.

\begin{figure}
\centering 
\includegraphics[scale=0.7]{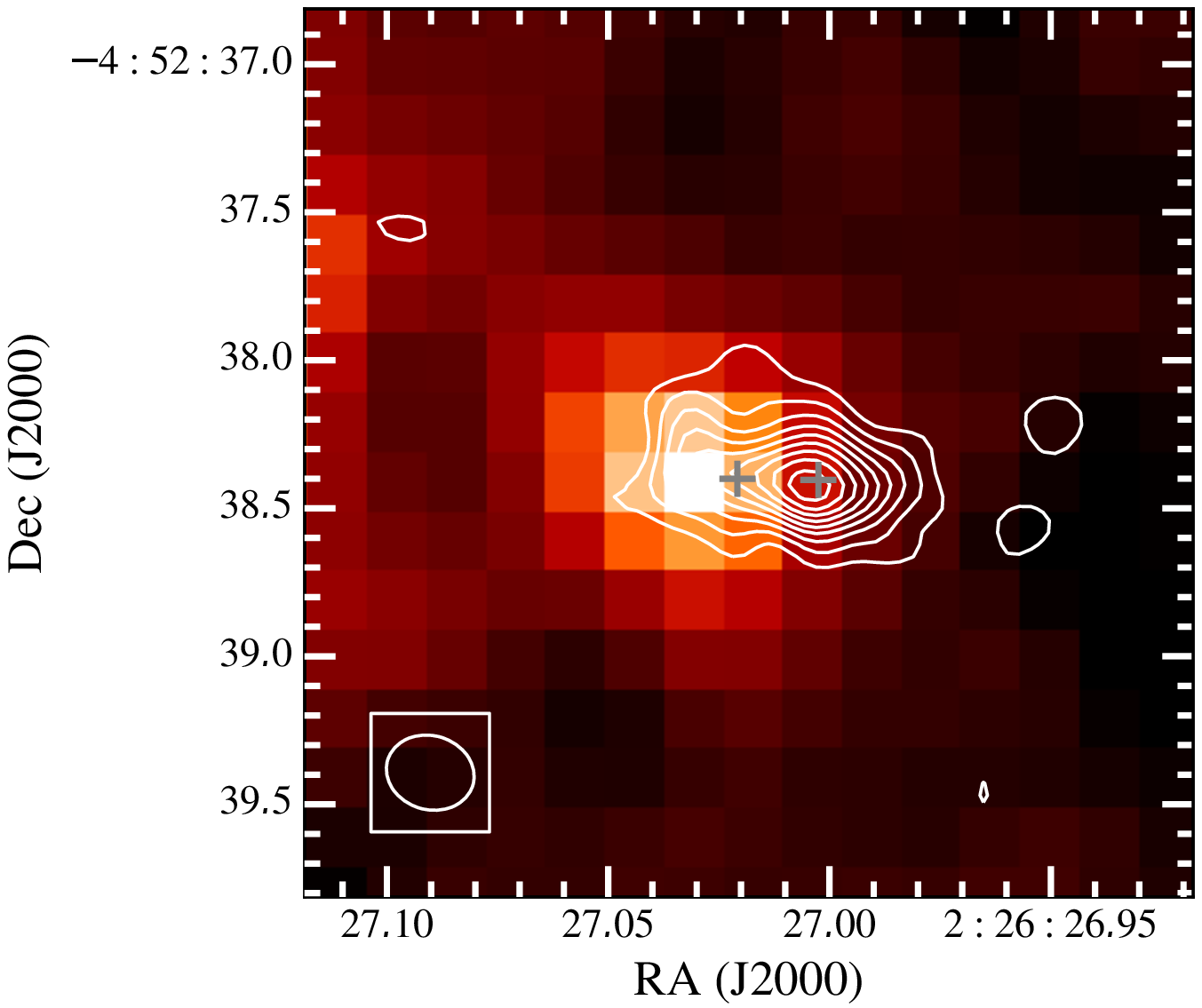}
\caption{Contours of \cii moment zero map overlaid on top of near-IR image. 
Contours begin at $\pm2\sigma$, where $1\sigma= 0.02\,$Jy\,beam$^{-1}$\,km\,s$^{-1}$, 
and are in steps of $1\sigma$. Crosses show locations of core and tail. The ALMA
synthesized beam was $0.30''\times0.25''$, with major axis position angle = $74^{\circ}$ 
(shown in lower-left). The near-IR image is an average of the Z and Y band images 
from the ESO VISTA VIDEO survey data release 4 \citep{jarv13} and has spatial resolution of $0.8''$.}
\label{fig:ciinir}
\end{figure}

Figure \ref{fig:ciinir} shows the velocity integrated (moment zero) \cii
emission contours overlaid on the ground-based near-IR image.  A 2D
Gaussian fit of the \cii emission returns a deconvolved size of
$(0.65\pm0.08)''\times(0.30\pm0.04)''$ at $(81\pm6)^{\circ}$, an
integrated flux density of $0.70\pm0.08$\,Jy\,km\,s$^{-1}$, and a peak
flux density of $0.19\pm0.02$\,Jy\,beam$^{-1}$\,km\,s$^{-1}$.  However
the \cii emission is non-Gaussian in shape due to the multiple
components already discussed. The centre of the near-IR emission is
located close to the tail position, not at the peak of the core, as
previously discussed by \cite{will15}. The spatial resolution of the
near-IR image is insufficient to determine the fraction of flux coming
from each region.

\begin{figure}
\centering 
\includegraphics[scale=0.4]{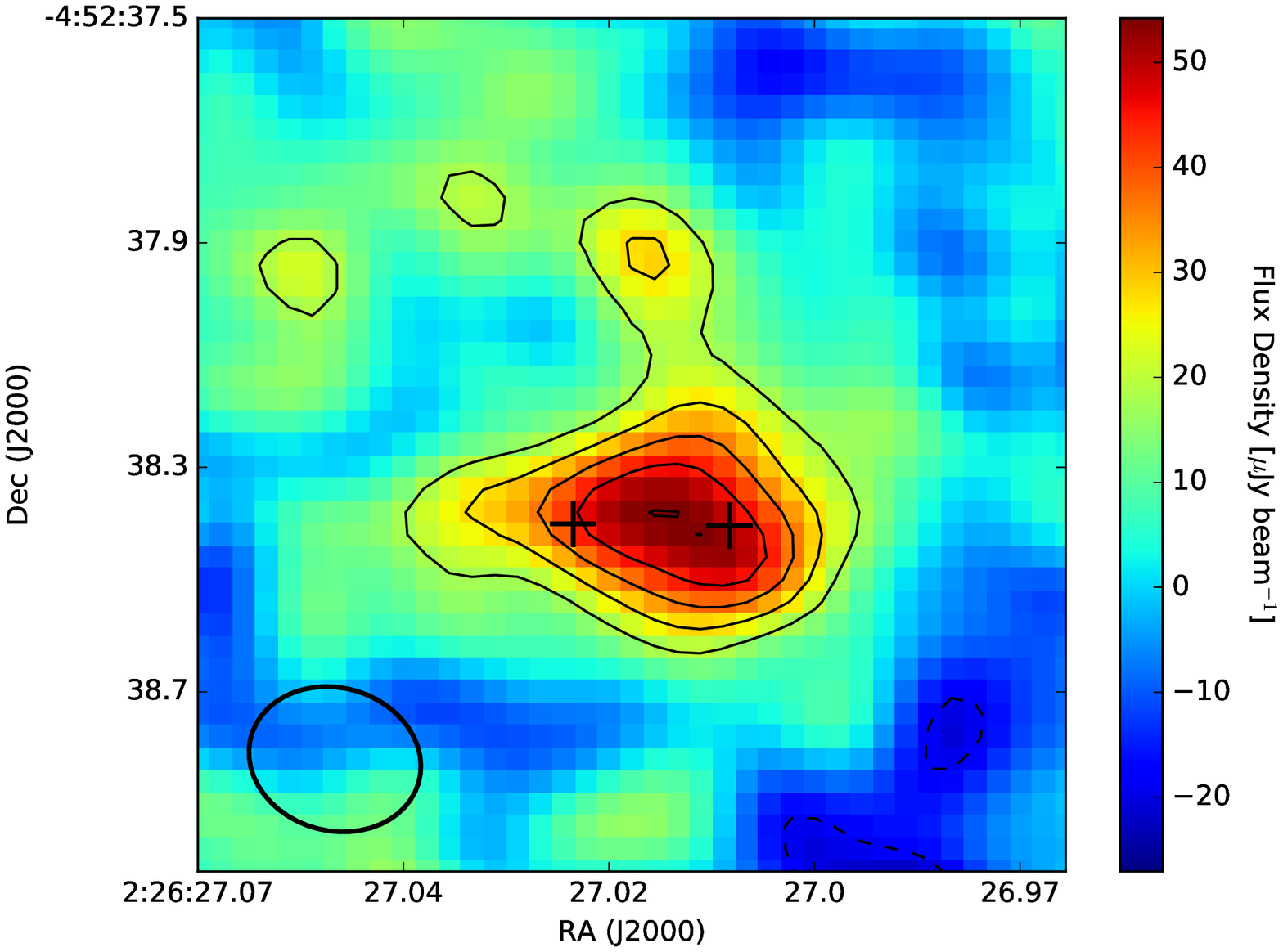}
\caption{Continuum image of WMH5. Contours begin at $\pm2\sigma$, where $1\sigma= 9\,\mu$Jy\,beam$^{-1}$, and are in steps of $1\sigma$. Crosses show locations of core and tail. The dark ellipse to the lower left shows the synthesized beam: $0.30''\times0.25''$, with major axis position angle = $74^{\circ}$.}
\label{fig:continuum}
\end{figure}

Figure \ref{fig:continuum} shows the $\sim160\,\mu$m continuum
emission. A Gaussian fit yields a deconvolved size of
$(0.5\pm0.1)''\times(0.3\pm0.1)''$ at $96\pm67^{\circ}$, a peak flux
density of $53\pm9\,\mu$Jy\,beam$^{-1}$, an integrated flux
density of $0.15\pm0.03\,$mJy, and a central position of 2h26m27.0099s $-4^{\circ}52'38.3770''$. The morphology is obviously different
from the beam shape, with an extension to the east and a $3\sigma$
extension to the north. The peak of the continuum is located between
the \cii core and tail positions, but since this emission is a blend of the core and tail components, this is to be expected.

\section{Analysis}

\subsection{\cii Luminosities \& SFRs}

\begin{figure}
\centering 
\includegraphics[scale=0.4]{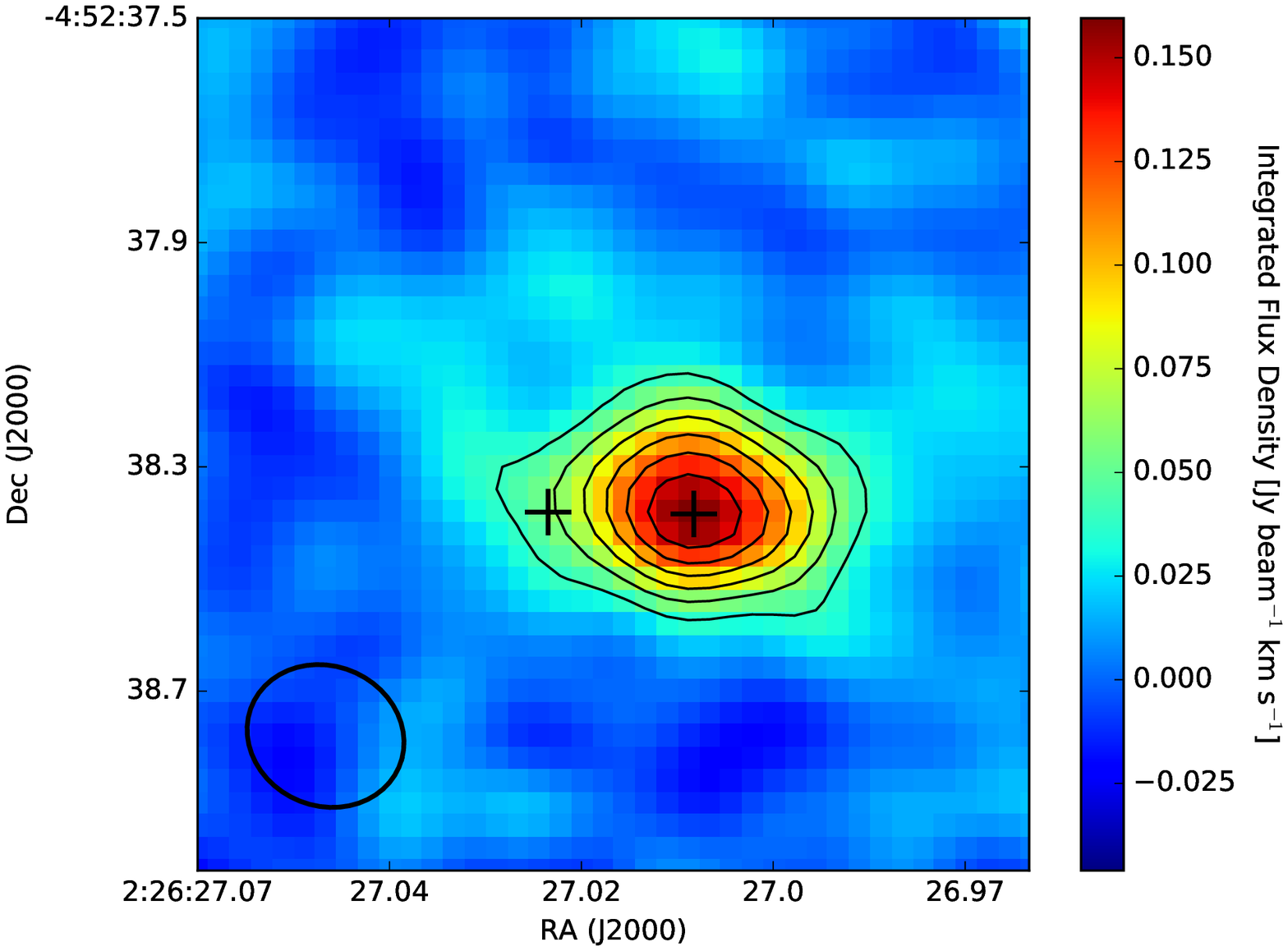}
\includegraphics[scale=0.4]{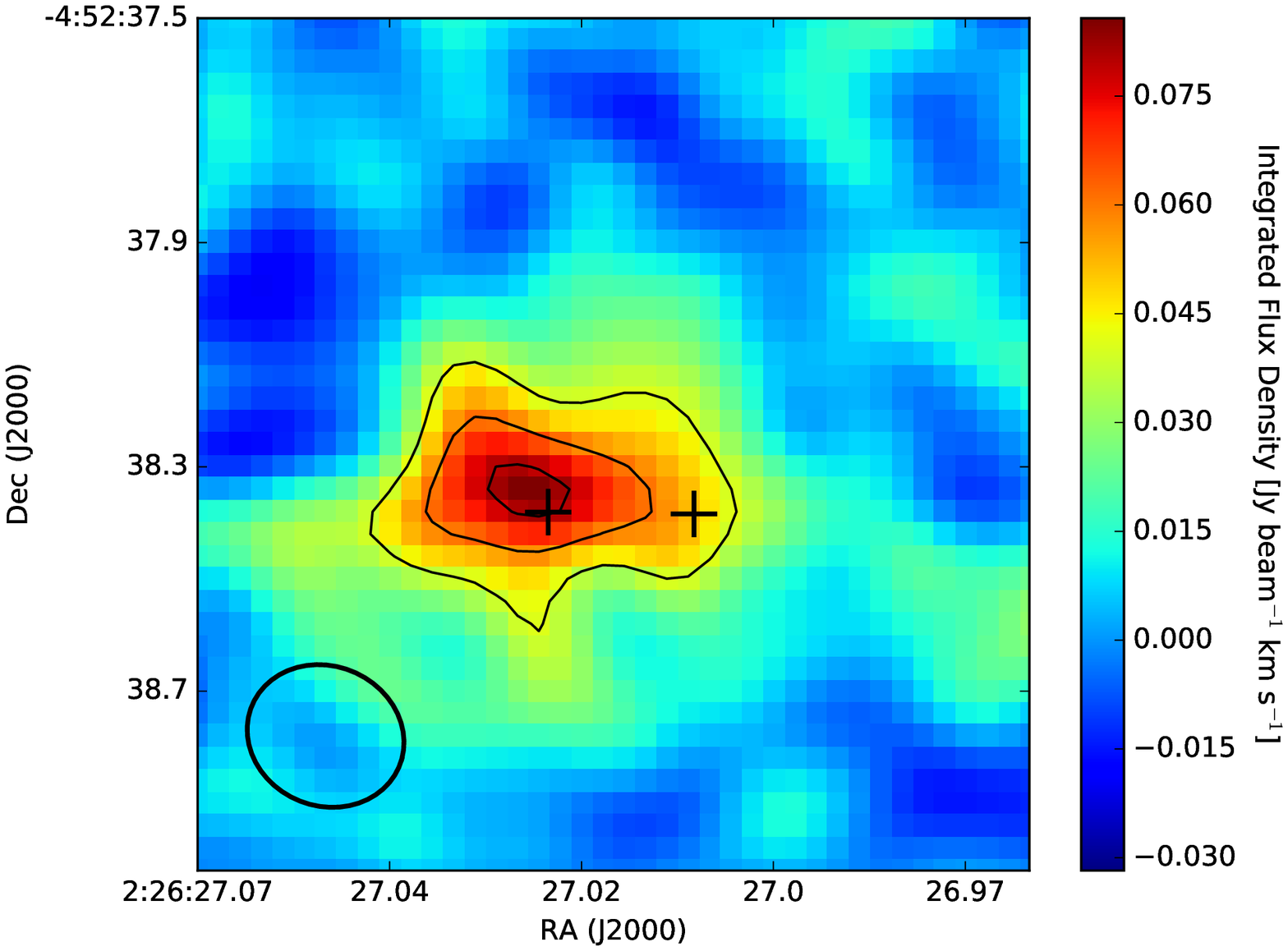}
\caption{Moment zero maps integrated over velocity ranges dominated by core galaxy (LEFT) 
and sub-galaxy (RIGHT) emission. Contours 
begin at $\pm4\sigma$, where $1\sigma= 0.01\,$Jy\,beam$^{-1}$\,km\,s$^{-1}$, and 
are in steps of $2\sigma$. Note the color scales of the maps differ by a factor of two. 
Crosses denote positions of maximum core and tail emission.  The dark ellipse to the lower left shows the synthesized beam: $0.30''\times0.25''$, with major axis position angle = $74^{\circ}$.}
\label{fig:twom0}
\end{figure}

We first analyze the line emission, integrated over the velocity
ranges corresponding predominantly to the core and sub-galaxies.  The
left side of Figure \ref{fig:twom0} shows the \cii emission
integrated over velocity ranges corresponding to the core galaxy (169
to -77\,km\,s$^{-1}$), while the right hand figure shows the \cii emission
integrated over the velocity
range of the sub-galaxies (-94 to -305\,km\,s$^{-1}$). We then
fit Gaussian models to the surface brightness distributions. 
We note that the spatial/spectral separation of the core and
sub-galaxies is not completely clean, so the flux densities of the 
sub-galaxy complex may be overestimated.

The core moment zero image shows a higher peak, but is spatially
more compact, with a deconvolved size of $(0.32\pm0.03)''\times(0.14\pm0.01)''$, and a velocity
integrated flux density of $0.32\pm0.03$\,Jy\,km\,s$^{-1}$.
The moment zero image of the sub-galaxies shows a larger spatial size, 
extending toward the core galaxy. The fit size is $(0.6\pm0.1)''\times(0.38\pm0.07)''$
and the velocity integrated flux density is
$0.30\pm0.04$\,Jy\,km\,s$^{-1}$. 
These \cii line fluxes imply \cii luminosities of:
$L_{\rm [C\,II],CORE}=(3.2\pm0.3)\times10^8\,L_{\odot}$ and
$L_{\rm [C\,II],SG}=(3.0\pm0.4)\times10^8\,L_{\odot}$.

While the combined flux densities of the tail components in Table \ref{north1} are considerably less than that of the core,
these luminosities are comparable. This is reasonable, since these luminosities are derived from 
flux densities integrated over the entire emission area of each source, while the values in Table \ref{north1}
represent only a single pixel. Since the tail emission is considerably more extended, its flux density at a given point
will be less than that of the concentrated core, but the sum of each component may be comparable. 

Using these luminosities, we can estimate the star formation rates,
assuming the relationship given in equation 8 of \citet{vall15} and a metallicity. 
Note that this relationship is based on the results of a cosmological simulation, where discrete values of metallicity (\textit{i.e.}, 1.0, 0.2, and 0.05\,Z$_{\odot}$) were assumed. This coarse sampling creates an inherent uncertainty in the relationship, which is exacerbated by additional assumptions (\textit{e.g.}, a Kennicutt-Schmidt power law index).  Assuming a solar metallicity, 
the implied star formation rates for the core and sub-galaxies are 
$\sim20\,M_{\odot}$ year$^{-1}$. If
we assume a metallicity of $0.1\,Z_{\odot}$, these values increase to 
$\sim80\,M_\odot$\,year$^{-1}$.

Equation 8 of Vallini et al. may also be used to solve for the metallicity of an object, given its SFR and \cii luminosity. To find the total SFR of WMH5, we add SFR$_{SED}$ and SFR$_{FIR}$ of \citet{will15}, yielding $76\pm11$\,M$_{\odot}$\,year$^{-1}$. This addition is due to the fact that the core galaxy (their galaxy `A') dominates the observed mm (rest-frame FIR) continuum emission, while the tail galaxies (their galaxy `B') is the source of the observed NIR (rest-frame UV) emission. Since the sources of the mm and NIR emission are spatially separated, and the constructed SED is mainly shaped by the UV data, the addition of the two SFRs is required to find the total value for the galaxy complex.

We use the total SFR and our total \cii luminosity of ($7.0\pm0.8)\times10^8\,L_{\odot}$, yielding $\sim0.2\,Z_{\odot}$ for the system as a whole. However, due to the discrete metallicities probed by Vallini et al. (see their Figure 8), we only state a metallicity limit of $0.1-1.0\,Z_{\odot}$.

\subsection{\cii-Continuum Ratio}

At this high resolution, we are able to analyze separately the
continuum and line surface brightness at the position of the core and
sub-galaxies.  The continuum surface brightnesses at the core and tail
positions are $(49\pm9)\,\mu$Jy\,beam$^{-1}$ and
$(42\pm9)\,\mu$Jy\,beam$^{-1}$, respectively.

For the line surface brightness, we use the spectral fitting results
shown in Table \ref{north1}. The core \cii velocity integrated
surface brightness is $0.21\pm0.01$\,Jy\,beam$^{-1}$\,km\,s$^{-1}$.  For the
tail position, the sum of the emission from the two sub-galaxies and
the diffuse component gives:
$0.15\pm0.08$\,Jy\,beam$^{-1}$\,km\,s$^{-1}$. The line to continuum ratio for the core is then: 
$(4.3\pm0.8)\times10^3$\,km\,s$^{-1}$. The value for the tail is
$(4\pm2)\times10^3$\,km\,s$^{-1}$.

We compare these ratios to observations of other high redshift
galaxies. In the sample of \citet{capa15}, which consisted of
$z\sim5.5$ main sequence galaxies observed in \cii, four
galaxies were detected in rest-frame $\sim158\,\mu$m continuum
emission. These four give ratios of $(6\pm2)\times10^3$\,km\,s$^{-1}$
(HZ4)\footnote{In Extended
Table 1 of \citet{capa15}, the \cii flux of HZ4 should be
$1.1\pm0.2$\,Jy\,km\,s$^{-1}$}, $(1.4\pm0.4)\times10^4$\,km\,s$^{-1}$ (HZ6),
$(3.8\pm0.3)\times10^3$\,km\,s$^{-1}$ (HZ9), and
$(1.25\pm0.08)\times10^3$\,km\,s$^{-1}$ (HZ10). Thus, the ratios of both segments of
WMH5 fall within the scatter of the Capak et al. sample. \citet{will15} show that the system as a whole has a \cii line luminosity to FIR continuum luminosity ratio $\sim 5\times 10^{-3}$, which is larger than low redshift star forming galaxies by a factor of a few. They point out that such a large ratio: `...suggests extended star formation with low metallicity and an intense radiation field.' 

\begin{figure}
\centering 
\includegraphics[scale=0.4]{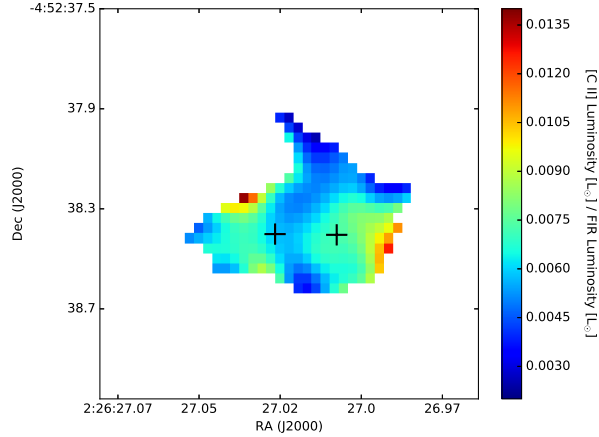}
\caption{The ratio of line emission (i.e., intensity of the \cii
moment zero map) to continuum emission over the WMH5 field. Only $1.5\sigma$
emission in both the continuum and \cii moment zero maps was considered. Crosses correspond to locations of core
(west) and tail (east) emission.}
\label{fig:ratio}
\end{figure}

In addition to these two discrete locations, we may examine the 
ratio of \cii to FIR luminosities over the source as a whole 
(see Figure \ref{fig:ratio}). The integrated \cii flux density
was converted to a luminosity using equation 1 of \citet{solo92}.
A greybody spectrum  identical to that used by \citet{will15} was adopted
to convert our continuum flux density to a FIR luminosity (i.e., $T_{\rm dust}$=30\,K, 
$\beta=1.6$, and a FIR range of $\lambda=42.5$ to 122.5\,$\mu$m). Note that we assume that
these parameters are uniform across the source. 

This ratio plot shows that while the core and far-east tail areas behave similarly,
there is a gap of low $L_{\rm [C\,II]}/L_{\rm FIR}$ between them. 
A similar area of low \cii emission is observed in the simulations of \citet{pall17}. In their Section 3.3.3, they reason that this `middle ground' contains diffuse (low $n$) molecular clouds, which only weakly emit \cii ($L_{\rm[C\,II]}\propto n$). In addition, the relatively high temperature of the CMB at $z\sim6$ ($\sim19$\,K) diminishes the \cii luminosity. 

\subsection{Velocity Behavior}
\begin{figure}
\centering 
\includegraphics[scale=0.5]{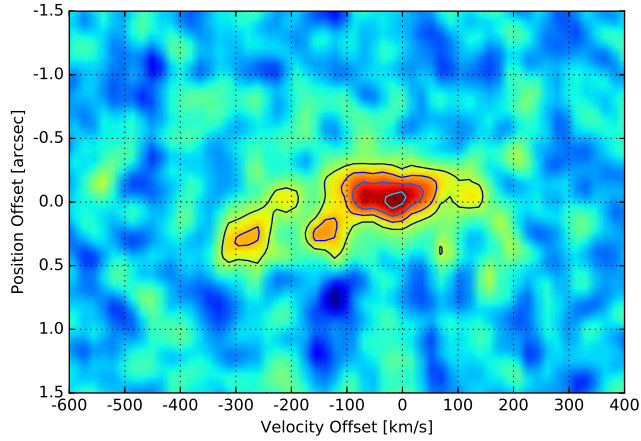}
\caption{Position-velocity diagram of WMH5, taken for $1.5''$ from a central position of 2h26m27.006s, -4$^{\circ}$52$'$38.420$''$, at an angle of $80^{\circ}$. Contours are displayed for (2,3,4)$\times0.17\,$mJy\,beam$^{-1}$.}
\label{fig:six}
\end{figure}

Figure \ref{fig:six} is a position-velocity (PV) plot.  Again, in
addition to the core emission at zero offset, two distinct tail
components are evident, plus diffuse emission throughout. The spatial
offset of the two tail components from the core is $0.24''$ in
projection, or about 1.4\,kpc.

This PV plot is similar to Figure 6 of \citet{will15}, although at higher 
spatial resolution. While they detected SG-1 (their component `B'),
we are able to separate their component `A' into the core galaxy and SG-2.
Both subgalaxies are clearly distinct from the core galaxy, and show similar positional offsets.

Our best-fit velocities (Table \ref{north1}) and those of \citet{will15} (their Figure 1)
are also in agreement, with component `A' ($v_{[C\,II]}=0\pm13\,$km\,s$^{-1}$) and 
our `core' ($v_{[C\,II]}=-1\pm5\,$km\,s$^{-1}$) in agreement. In addition, component `B'
($v_{[C\,II]}=-238\pm13\,$km\,s$^{-1}$) and SG-1 ($v_{[C\,II]}=-249\pm5\,$km\,s$^{-1}$) agree.

\begin{figure}
\centering 
\includegraphics[scale=0.5]{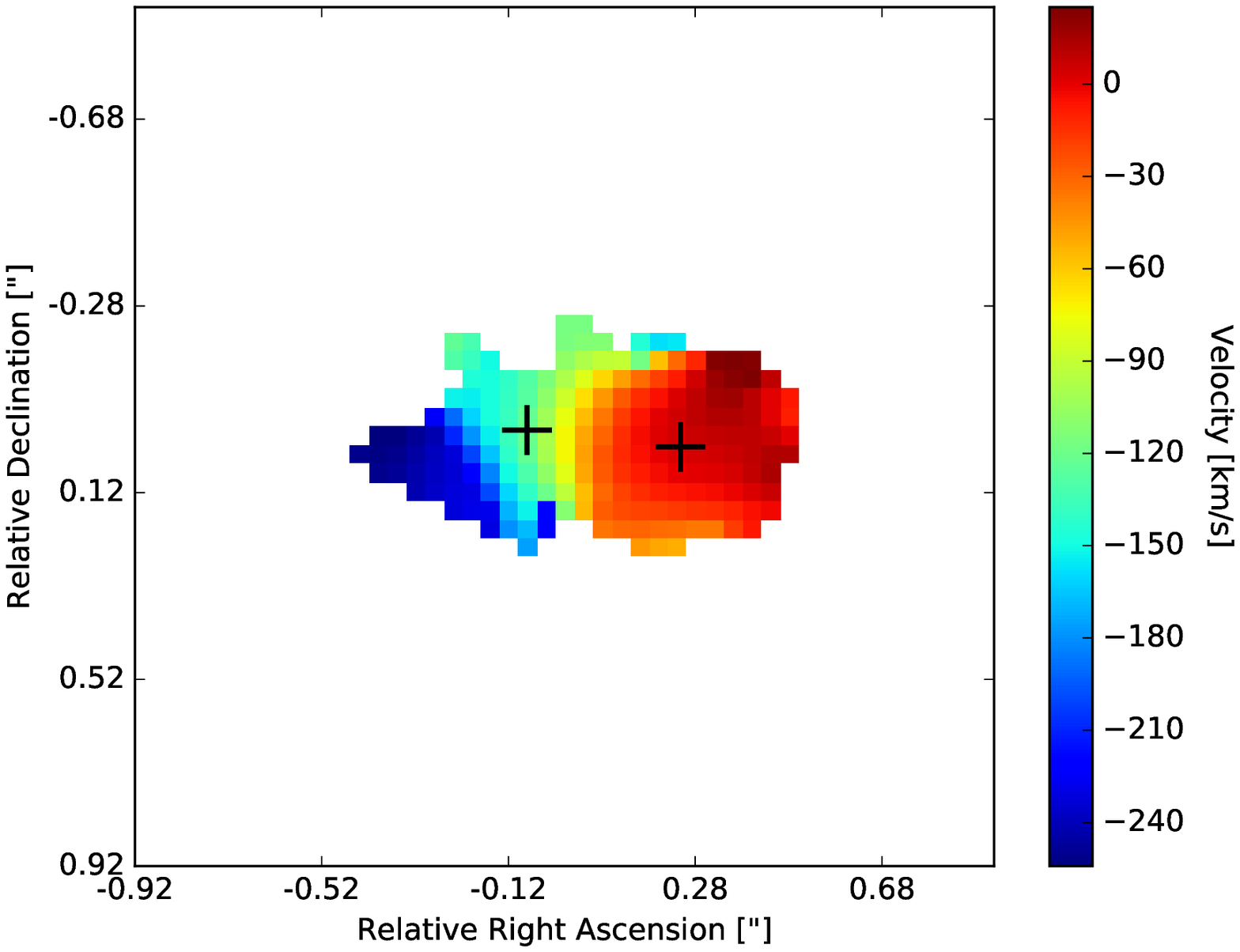}
\caption{Moment 1 (velocity field) map of WMH5. Zero velocity is defined at the \cii rest 
frequency shifted to $z=6.0695$.}
\label{fig:seven}
\end{figure}

While the velocity of the tail components is multivalued, we 
may create a velocity weighted (moment 1) image of the core 
galaxy. The zero of velocity here is assumed to be the \cii 
emission redshifted to $z=6.0695$. Figure \ref{fig:seven} shows 
that the core galaxy is not highly reminiscent of either a 
solid-body rotator or a classic rotating disk, but does show 
an identifiable velocity gradient.

\subsection{Dust Mass}
By using equation 1 of \citet{bian13}, our $\sim160\,\mu$m continuum flux density ($0.15\pm0.03\,$mJy), 
and the dust temperature and emissivity index adopted by \citet{will15}, we find a dust mass of 
$(4.0\pm0.8)\times10^8\,M_{\odot}$. For the dust absorption coefficient, we use $\kappa_{\nu}=\kappa_o\left( \nu /\nu_o \right)^{\beta}$
 of \citet{wang11}, with $\kappa_o=1.88$\,m$^2$\,kg$^{-1}$ at $\nu_o=2.4$THz.
 
However, due to the weak constraints that we have on the assumed SED features, we additionally assume 
ranges for dust temperature and emissivity index of $T_{\rm dust}=25-100$\,K 
and $\beta=1.2-2.0$ (\citealt{capa15,neel17}). This conservative range yields a range of 
possible dust masses of the WMH5 system of $M_{\rm dust}=10^{7.4-8.9}\,M_{\odot}$.

This range may be compared to other high redshift galaxies with similar SFRs (i.e., $\sim10^{1-2}$\,$M_{\odot}$\,year$^{-1}$).
The gravitationally lensed object A2744-YD4 ($z=8.38$; \citealt{lapo17}) exhibits a SFR$\sim20$\,$M_{\odot}$\,year$^{-1}$, 
but a small dust mass of $\sim10^{6.8}$\,$M_{\odot}$. This relatively small amount of dust may be simply explained by the 
different ages of each object (WMH5 is observed 300\,Myr later). This reflects differing amounts of past star formation activity that each object has undergone.

If the same assumptions are made to transform the $\sim160$\,$\mu$m continuum flux densities of the \citet{capa15} objects to dust masses, we find a range of $M_{\rm D}=10^{7.4-9.9}$\,M$_{\odot}$, which are similar to our estimates for WMH5.

\section{Discussion}

The merger interpretation emerging from the present observations 
is supported by recent AMR, zoomed simulations of high 
redshift galaxies with properties similar to WMH5. \citet{pall17} 
studied in detail \textit{Dahlia}, a typical $z\approx 6$ LBG 
with a SFR $\approx 100\,M_{\odot} \mathrm{yr}^{-1}$, and an effective 
stellar radius of $\sim 0.6$\,kpc. This model galaxy has 14 satellites distributed 
over a distance of 100\,kpc from it. The six largest ones have a dark matter halo mass 
in the range $M_h = 2.5-12 \times 10^9 M_\odot$, three among these are located within 3\,kpc from 
Dahlia's center (whereas our sub-galaxies are $<6$\,kpc distant).
The two most prominent among these 
satellites (at a distance of 12 and 15 kpc from the center) 
sustain considerable star formation rates of 30 and $10\,M_{\odot} 
\mathrm{yr}^{-1}$, and have a stellar mass of $M_*=6 \times 10^9 M_\odot$ 
and $M_*=2 \times 10^9 M_\odot$. Their star formation history is relatively bursty 
and increases with time. They are also the oldest among the satellites, with an age 
at $z=6$ of 350 Myr.  Compared to \textit{Dahlia} ($M_*=1.6\times10^{10}$) their stellar 
mass is 38\% and 12\% smaller, respectively. The satellites are mostly aligned along 
three main filaments connecting to the center (see Fig. 4 of \citealt{pall17}); they are
also much more compact than \textit{Dahlia}: the largest satellite within 3 kpc has an
effective radius of 0.2\,kpc, i.e. about 1/3 of that the central galaxy.

Interestingly, the three nearby satellites are clearly visible in the
simulated \cii maps, thus supporting the analogy with our
observations. The satellites move at $\sim 80$\,km\,s$^{-1}$,
i.e. about half the virial velocity of \textit{Dahlia}. They have 
also very low metallicity, due to their relatively young age ($\le
100$\,Myr). Compared to \textit{Dahlia}, which has a metallicity
equal to $0.5\,Z_{\odot}$, the nearby satellites barely reach $Z =
10^{-2}\,Z_{\odot}$; the two distant ones instead have $Z=0.03\,Z_\odot$ and 
$Z=0.1\,Z_\odot$ . If the dust content scales with metallicity, this
should result in a substantially lower extinction. Thus, the
comparison with theoretical models supports the idea that WMH5
features the build-up of a galaxy occurring via infall and merging of
a number of star-forming satellite sub-galaxies.

Of course, we have to be aware that the comparison with observations is here 
performed only against a single, although very similar, simulated system. Thus, 
it is probably unsafe to push the comparison beyond the present analysis. 
Before drawing more robust conclusions one would need to extend the study 
to a larger sample of simulated galaxies which could inform us on the frequency 
of multiple/merger systems expected at these early redshifts. Nevertheless, the 
emerging scenario is one in which an intriguing resemblance between theoretical 
expectations and the data analysis certainly exists.

Our geometry is also akin to that found in \citet{katz16} (their Figure 20), in that
we have a `filament' of gas falling into a galaxy, with spatially separated
\cii and UV/Ly$\alpha$ emission. Katz et al. interpret this geometry as being caused by the infalling galaxy causing new star formation
at the interface of the filament and galaxy, while introducing relatively pristine (low metallicity) gas.
This low-$Z$ environment is weak in \cii emission, but high in UV/Ly$\alpha$.
Deeper inside the host galaxy, the environment is more processed (higher metallicity), so while the UV/Ly$\alpha$
emission is obscured, \cii emission may escape. 

\section{Conclusion}

New ALMA observations of the dust continuum and \cii 158\,$\mu$m
emission from the $z=6.0695$ LBG WMH5 show a compact main
galaxy in dust and \cii emission, with two, what we call,
`sub-galaxies' located along a filamentary tail extending about 5\,kpc
(in projection) to the east. These sub-galaxies are distinct in
velocity, with narrow velocity dispersions of about 80\,km\,s$^{-1}$,
and the tail joins smoothly into the main galaxy velocity field. The
sub-galaxies themselves are extended east-west by about $0.6''$ in
individual channel images.


We conjecture that the WMH5 system represents the early formation of a
galaxy through the accretion of smaller satellite galaxies embedded in
a smoother gas distribution, along a filamentary structure.  While we
cannot rule-out outflow, we find that notion less likely, given the
narrow velocity dispersion of the individual sub-galaxies. Such narrow
velocity dispersion is more characteristic of gravitational or tidal
features than strong outflows. Likewise, the significant dust masses
and star formation rates suggest individual sub-galaxies, as might be
expected in a coalescing galaxy system.

Our filamentary interpretation assumes a three-dimensional structure
that we are unable to directly observe. However, the continuity of the
velocity field from the core to each sub-galaxy, the small velocity
widths of each sub-galaxy, the intrinsic E-W extension in channel
images of the sub-galaxies themselves, and the similarity of these
results to those found using three-dimensional cosmological
simulations all provide circumstantial evidence supporting this
conclusion, although the term filamentary remains presumptive.

These observations highlight the power of ALMA to image early galaxy
formation. The results support the idea that early galaxies can be
enriched in heavy elements and dust due to their very early history of
star formation, including enrichment of the material that accretes
onto main galaxy.  The new Band 5 system of ALMA is currently
being commissioned (163\,GHZ to 211\,GHz). This band will open the $z = 8$
to 11 range for high resolution, sensitive observations of the
\cii 158\,$\mu$m line.

\vspace{5mm}

This  paper  makes  use  of  the  following  ALMA  data:
ADS/JAO.ALMA \#2013.1.00815.S and \#2015.1.00834.S. ALMA is a partnership of ESO
(representing its member states), NSF (USA) and NINS (Japan),
together with NRC (Canada) and NSC and ASIAA (Taiwan), in co-
operation with the Republic of Chile. The Joint ALMA Observatory
is operated by ESO, AUI/NRAO and NAOJ. GCJ is grateful for support from NRAO through the Grote Reber Doctoral Fellowship Program. AF acknowledges support from the ERC Advanced Grant INTERSTELLAR H2020/740120.

\clearpage
\newpage

\end{document}